\documentclass[a4paper]{article}

\usepackage{IEEE}

\title{Convolutional Variational Autoencoders for Audio Feature Representation in Speech Recognition Systems}
\name{Olga Yakovenko$^{1,2}$, Ivan Bondarenko$^1$}
\address{
  $^1$NSU, Russia\\
  $^2$Zolotaya Korona, Russia}
\email{o.iakovenko@g.nsu.ru, i.yu.bondarenko@gmail.com}

\begin{document}

\maketitle
\begin{abstract}
For many Automatic Speech Recognition (ASR) tasks audio features as spectrograms show better results than Mel-frequency Cepstral Coefficients (MFCC), but in practice they are hard to use due to a complex dimensionality of a feature space. The following paper presents an alternative approach towards generating compressed spectrogram representation, based on Convolutional Variational Autoencoders (VAE). A Convolutional VAE model was trained on a subsample of the LibriSpeech dataset to reconstruct short fragments of audio spectrograms (25 ms) from a 13-dimensional embedding. The trained model for a 40-dimensional (300 ms) embedding was used to generate features for corpus of spoken commands on the GoogleSpeechCommands dataset. Using the generated features an ASR system was built and compared to the model with MFCC features.
\end{abstract}

\section{Introduction}

Automatic recognition of the spoken language has already became a part of a daily life for many people in the modern world. Although algorithms for automatic speech recognition have progressed greatly throughout the last years, most of the applications still utilize a basic set of features – Mel-frequency cepstral coefficients. These kind of features are processed rapidly and can produce good results, but recent research \cite{Amodei2016} has proven that raw FFT spectrograms give better accuracy on ASR tasks when combined with Deep Neural Networks (DNN) for feature-extracting.

The drawback of using spectrograms as the inputs to a recognition system is that a lot of computational power has to be used to process a single audio. With the increase of length of the audio increases the time spent computing an optimal path from first to last sound. Embeddings, generated by VAE, can be used as a compressed version of the general FFT spectrogram features for both traditional Gaussian Mixture Models and modern DNNs. Moreover, this approach can be used for dimensionality reduction of input data \cite{Dai2017} that can subsequently be used for visualization of audio data or for reduction of occupied space on a hard drive.

Traditionally VAE is used as a generative model, similar to Generative Adversarial Networks (GAN). For the past several years there has been going a lot of research on the topic of audio generation with VAEs, including the fascinating DeepMind’s VQ-VAE \cite{Oord2017}, which is a successor of WaveNet \cite{Oord2018}. The main features of those two systems is that having the train samples of some audio clips the system learns the probability distributions over observed features, making it possible to generate new samples by selecting other values from these distributions.

VAE was found out to be a good tool for dimensionality reduction using the bottleneck features of the hidden layer. For example, authors of \cite{Dai2017} prove in their paper that VAE is a natural evolution of the robust PCA algorithm. Basic Autoencoders (AE) are also known to be able to reduce the dimensionality of the data, but the resulting embedding does not serve as a good generalization, because the hidden representations of the intermediate layer are poorly structured due to non-probabilistic nature of AE.

Over the past few years, there has also been attempts to perform construction of audio embeddings. The main goal of this approach is to be able to compare many utterances to some etalon utterance or to compare different utterances against each other and present all of them in some feature space. This approach could be found to be used in the spoken term recognition by example \cite{Zhu2018}, for speaker identification or verification using Siamese Neural Networks \cite{Milde2018} and for semantic embedding construction based on context based on Recurrent Neural Networks (RNN) and skipgram/cbow architectures \cite{Chung2018}.

It is important to notice, that most of the presented approaches use recurrent structures for analyzing variable-length audio data. Recurrent structures enable good quality of automatically determining features, relevant to the task, having approximately same length. The drawback here is that if the system was trained on samples of length from 4 second to 10 seconds, it will produce unexpected results for samples much longer than 10 seconds. Researcher is forced to either adapt his data to fit the model, or to build his own model that would fit his data. These drawbacks can be avoided by using fixed-length features, such as MFCC or VAE-based features, presented in the following paper.

Thus, in this paper an approach towards feature generation for audio data is presented. The first part is dedicated to the feature generation itself and the ability to reconstruct the signal from the resulting embedding, while the second part describes the experiments with the embeddings for an ASR task.

\section{Traditional audio features for ASR/TTS tasks}

Audio wave is a sequence of positive and negative values, the values represent the amplitudes of the fluctuation of a sound device. The frequency and the amount of the fluctuation defines different sounds.

\subsection{Raw wave}

Sometimes raw signal is used for speech recognition or generation tasks. The only preprocessing that the sound may undergo is discretization of an analog sound when it is being recorded. It is possible to analyse a discrete signal, if the task itself is not too complex (e.g. recognition of loud noises, of sound patterns) but also if the system consists of an end-to-end structure that besides solving the main tasks is also capable to detect frequencies in a signal. Some of those end-to-end architectures include 1D Convolutions or PixelCNNs.

\begin{figure}[t]
  \centering
  \includegraphics[width=\linewidth]{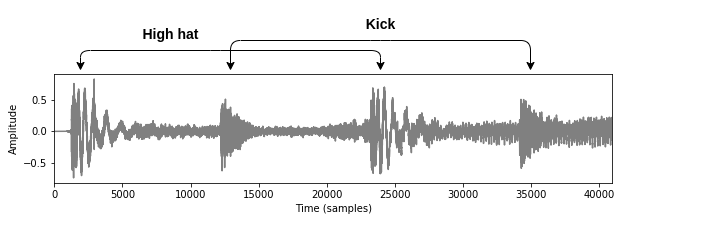}
  \caption{Amplitude audio representation.}
  \label{fig:audio_wave}
\end{figure}

\subsection{Spectrogram}

A traditional approach towards deep analysis of speech and sounds is mostly realized through spectrograms using a Short-Time Fourier Transform (STFT). Spectrograms are quite useful for manual analysis: for example, for detection of formants in human's speech. Generally, spectrograms are very good means of learning information from audio data, since sounds, that are produced by the vocal cord and, more importantly, speech, are characterized by cooccurrence of different harmonics and their change through time. STFT is applied to a discrete signal, so the data is broken up into overlapping chunks. Each chunk is Fourier transformed, and the result is added to a matrix, which records magnitude and phase for each point in time and frequency. This can be expressed as:

\begin{equation}
\begin{gathered}
\mathbf {STFT} \{x[n]\}(m,\omega )\equiv X(m,\omega )= \\
\sum _{n=-\infty }^{\infty }x[n]w[n-m]e^{-j\omega n}
\label{stft}
\end{gathered}
\end{equation}

with signal $x[n]$ and window $w[n]$.

The magnitude squared of the STFT yields the spectrogram representation of the Power Spectral Density of the function:

\begin{equation}
\operatorname {spectrogram} \{x[n]\}(m,\omega )\equiv |X(m,\omega )|^{2}
\label{spectrogram}
\end{equation}

The described spectrogram was generated for the LibriSpeech dataset during the experiments with VAE.

One of the commonly used modifications of STFT spectrograms are mel-spectrograms. During this modification each of the time frames in a spectrogram are passed through series of triangular filters (mel-scaled filters). Mel filters are based on a hand-crafted mel-scale, which represents the extent to which humans can distinguish sounds of different frequencies.

\begin{figure}[t]
  \centering
  \includegraphics[width=\linewidth]{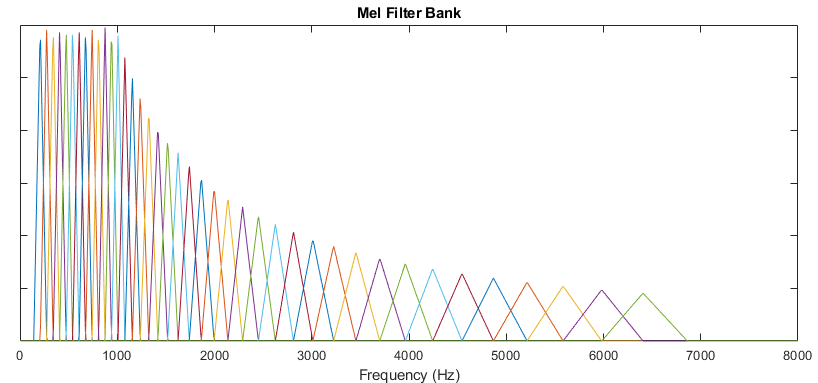}
  \caption{Mel-scaled filterbanks.}
  \label{fig:mel-filterbanks}
\end{figure}

All of the aforementioned approaches have led us to the traditional approach for creation of audio features: Mel-Frequency Cepstral Coefficients (MFCC).

\subsection{MFCC}

Once we have the mel-spectrograms, we can then proceed to the computation of the Mel-Frequency Cepstral Coefficient (MFCC). It consists of three simple steps:

\begin{enumerate}
\item Compute elementwise square of each value in the mel-spectrogram;
\item Compute elementwise logarithm of each element;
\item Apply discrete cosine transform to the list of mel log powers, as if it were a signal.
\end{enumerate}

The final result will also be a matrix, quite unreadable by a human eye, but surprisingly effective when used for both Speech to Text (STT) and Text to Speech (TTS). To this day MFCC features are used mostly in all speech-related technologies.

\begin{figure}[t]
  \centering
  \includegraphics[width=\linewidth]{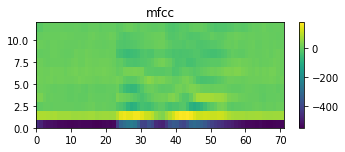}
  \caption{MFCC of the LibriSpeech sample 1779-142733-0000.wav.}
  \label{fig:mfcc_example}
\end{figure}

Mel-spectrogram and MFCC are means towards compressing audio data without erasing the information relevant to speech, since these features are further used in applications, connected to speech.

Here we determine the goal of this study: we believe that it is possible to compress audio in analogous way, but with the help of neural network, specifically, a Variational Autoencoder. We are going to cover the architecture in the next chapter.

\section{Audio VAE}

As was mentioned in the introduction, Variational Autoencoders can be successfully used not just as a generative network, but also as a dimensionality reduction tool. This is the one of the main reasons for selecting this architecture. The other reason for selecting a VAE for audio compression is the assumption that a VAE that is trained on a big clean corpus of speech will then be able to extract features from a spectrogram that are relevant only to the speech. Therefore, the encoder of the resulting VAE will be able to conduct denoising, selection of relevant frequencies and compression simultaneously. As an interesting additional feature, this kind of network may be used for generating sounds that sound close to the true signal, that way acting as a option for augmentation.

It is also important to specify that encoder and decoder of the VAE include convolutional layers instead of fully-connected layers to evade a big amount of parameters to optimize and improve the recognition of formants of various frequencies and scales.

\subsection{Architecture}

\begin{figure}[t]
  \centering
  \includegraphics[width=\linewidth]{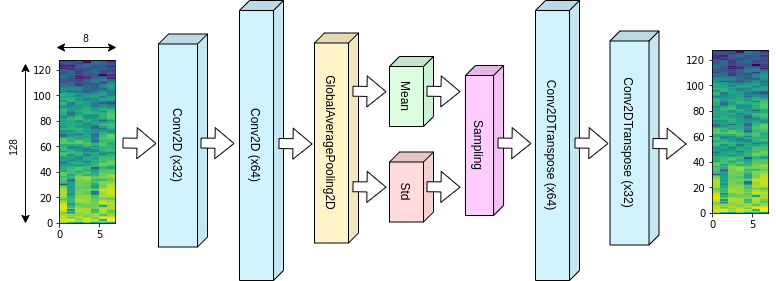}
  \caption{VAE architecture.}
  \label{fig:my_vae}
\end{figure}

VAE architecture was inspired by traditional VAE for image generation and segmentation. The VAE neural network is symmetric, where encoder and decoder parts include 2 convolution layers with kernel sizes 8 and stride 2. The encoder and decoder part are connected by a bottleneck which consists of 2D global average pooling layer, then followed by mean and variance layers of the VAE and finally the sampling layer (Figure~\ref{fig:my_vae}).

As for any other VAE, special loss function had to be defined that consists of two parts: reconstruction loss and regularisation. In our case reconstruction loss is Mean Squared Error (MSE) between the true values and predicted values and regularisation is estimated by Kullback–Leibler divergence between the distribution that is present in the data and the distribution that has been modeled by our encoder.

\begin{equation}
\begin{gathered}
  \mathcal {L}(\mathbf {\phi } ,\mathbf {\theta } ,\mathbf {x}, \mathbf {x'} )=\\
  \|\mathbf {x} -\mathbf {x'} \|^{2} + \\
  -0.0005 * D_{\mathrm {KL} }(q_{\phi }(\mathbf {h} |\mathbf {x} )\Vert p_{\theta }(\mathbf {h} ))
  \label{vae_loss}
\end{gathered}
\end{equation}

where $p_{\theta }(\mathbf {x} |\mathbf {h} )$ is the directed graphical model that we want to approximate, encoder is learning an approximation of this model $q_{\phi }(\mathbf {h} |\mathbf {x} )$ to the Gaussian multivatiate (in our case) distribution $\displaystyle p_{\theta }(\mathbf {h} |\mathbf {x} )$ where $\mathbf {\phi }$ and $\mathbf {\theta }$  denote the parameters of the encoder (recognition model) and decoder (generative model) respectively.

\subsection{Dataset}

Training of the system was carried out on a subset of the LibriSpeech dataset \cite{LibriSpeech}. LibriSpeech flac files were transformed to mono-channel wav, sampling rate 16 kHz, signed PCM little endian with bit rate 256 kB/s and bit depth 16. The subset was formed as follows:

\begin{itemize}
  \item Dataset was split into train and test with 2097 audiofiles representing train and 87 representing test.
  \item Train set included utterances from 1031 speakers, test set – 40 speakers.
  \item Spectrograms were calculated from the audio data.
  \item Each utterance was split into fragments of 0.025 sec with stride of 0.01 sec.
\end{itemize}

That left us with 1047736 samples for training and 30548 samples for testing. One input sample had the shape of 8 timestamps by 128 wave frequency stamps.

The choice for making such a small window (a small 25 ms window) was determined by different ASR systems, which use MFCC as their features. Later we want to compare the performance of an ASR system with VAE features and MFCC features, thus it is important to maintain the same conditions for feature generation for clear comparison.

\begin{figure}[t]
  \centering
  \includegraphics[width=\linewidth]{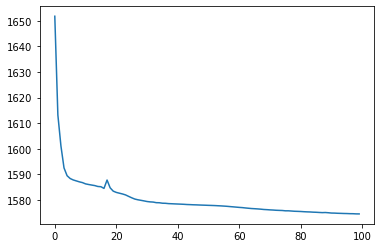}
  \caption{Train loss dynamics during training for 100 epochs.}
  \label{fig:train_loss}
\end{figure}

\begin{figure}[t]
  \centering
  \includegraphics[width=\linewidth]{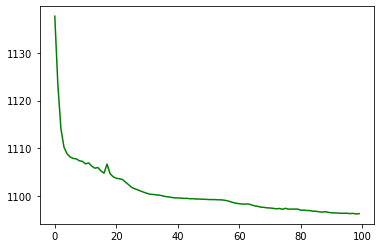}
  \caption{Test loss dynamics during training for 100 epochs.}
  \label{fig:test_loss}
\end{figure}

The dynamics for the training of the VAE audio features are shown on the Figure~\ref{fig:train_loss} for the train set and Figure~\ref{fig:test_loss} for the test set.

\subsection{Reconstruction}

\begin{figure}
\centering
\begin{minipage}{.10\textwidth}
  \centering
  \includegraphics[scale=0.5]{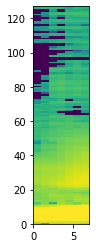}
  \label{fig:reconstruction_1}
\end{minipage}%
\begin{minipage}{.10\textwidth}
  \centering
  \includegraphics[scale=0.5]{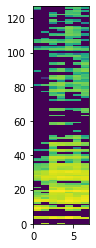}
  \label{fig:reconstruction_2}
\end{minipage}%
\begin{minipage}{.10\textwidth}
  \centering
  \includegraphics[scale=0.5]{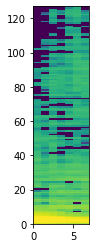}
  \label{fig:reconstruction_3}
\end{minipage}
\caption{Examples of reconstruction (lower) of the samples from the 13-dimensional hidden representation of the real sample (upper). The dark purple segments in the reconstructed pictures are the elements of matrix, that are equal to 0.}
\label{fig:reconstruction}
\end{figure}

The system described earlier was used for the encoding and reconstruction of the audio. Once we have trained the VAE, we can compress and decompress some samples from the dataset, some examples of the original and reconstructed samples can be seen in the Figure~\ref{fig:reconstruction}.

\begin{figure}
\centering
\begin{minipage}{.10\textwidth}
  \centering
  \includegraphics[scale=0.5]{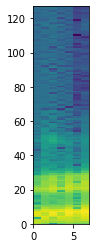}
  \label{fig:original_1}
\end{minipage}%
\begin{minipage}{.10\textwidth}
  \centering
  \includegraphics[scale=0.5]{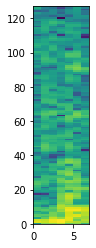}
  \label{fig:original_2}
\end{minipage}%
\begin{minipage}{.10\textwidth}
  \centering
  \includegraphics[scale=0.5]{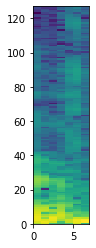}
  \label{fig:original_3}
\end{minipage}
\end{figure}

We can see from the pictures, that despite a big compression rate from 1024 elements of the spectrogram to just 13 elements of the latent vector, the information about the energised parts of the spectrum is still preserved. Moreover, the system seems to have learned the common localisation of the human speech (lower frequencies) and the way it can change over time.

\begin{figure}
\centering
  \includegraphics[scale=0.5]{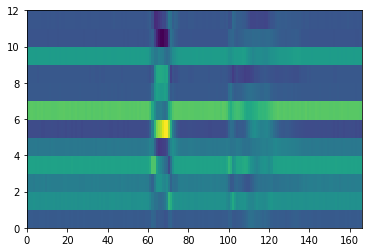}
\caption{Example of VAE encoding of the LibriSpeech sample 1779-142733-0000.wav.}
\label{fig:vae_example}
\end{figure}

The next step would be to visually compare the VAE features that we have achieved with MFCC features. MFCC for a file from the LibriSpeech dataset we have calculated earlier, see Figure~\ref{fig:mfcc_example}, and now we can see the analoguous features for the VAE generated encoding on the Figure~\ref{fig:vae_example}. It is evident from the pictures, that VAE has a better ability to code necessary segments of audio where human speech takes place, and seems to ignore the any kind of other sounds, that may sound on the background. On the other hand, MFCC features seem to detect some of the activity in the background, which may not reflect on ASR systems too well.

\subsection{Generation}

Since the trained encoder of the VAE is an approximation to a Gaussian multivariate distribution and the decoder is a directed graphical model, we can sample examples of spectrograms from the distribution and visualize them using our decoder.

\begin{figure}
\centering
\begin{minipage}{.10\textwidth}
  \centering
  \includegraphics[scale=0.5]{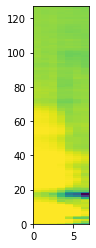}
  \label{fig:sample_3_1}
\end{minipage}%
\begin{minipage}{.10\textwidth}
  \centering
  \includegraphics[scale=0.5]{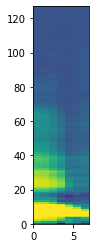}
  \label{fig:sample_3_2}
\end{minipage}%
\begin{minipage}{.10\textwidth}
  \centering
  \includegraphics[scale=0.5]{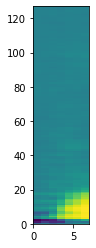}
  \label{fig:sample_3_3}
\end{minipage}%
\begin{minipage}{.10\textwidth}
  \centering
  \includegraphics[scale=0.5]{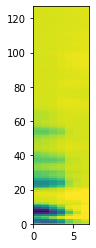}
  \label{fig:sample_3_4}
\end{minipage}
\caption{Samples from the multivariate (13 variables) Gaussian distribution, modeled by VAE, alternation of the 2-nd (top) and 5-th (bottom) variable, while the other are kept as mean values for the distribution.}
\label{fig:samples}
\end{figure}

For every component in the latent representation 4 evenly spaced numbers are chosen in the interval [-1, 1]. Then the resulting 4 vectors that differ by one value are transformed into spectrograms. Examples for some components are presented in Figure~\ref{fig:samples}.

\begin{figure}
\centering
\begin{minipage}{.10\textwidth}
  \centering
  \includegraphics[scale=0.5]{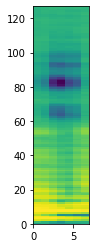}
  \label{fig:sample_1_1}
\end{minipage}%
\begin{minipage}{.10\textwidth}
  \centering
  \includegraphics[scale=0.5]{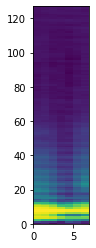}
  \label{fig:sample_1_2}
\end{minipage}%
\begin{minipage}{.10\textwidth}
  \centering
  \includegraphics[scale=0.5]{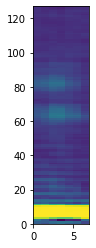}
  \label{fig:sample_1_3}
\end{minipage}%
\begin{minipage}{.10\textwidth}
  \centering
  \includegraphics[scale=0.5]{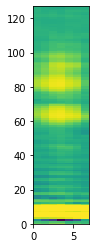}
  \label{fig:sample_1_4}
\end{minipage}
\end{figure}

It can be seen from the figures, that the VAE has learned how human speech may sound and has successfully encapsulated that knowledge in a multivariate Gaussian distribution.

\section{Experiments}
\label{experiments}

It is essential to compare developed approach with the existing. Therefore, experiments were carried out to evaluate the performance of resulting embeddings in a speech recognition task in comparison to MFCC features. Although the previous part was dedicated to reconstructing 25 ms feature frames, in this part 300 ms feature frames are regarded. The experiment with 25 ms frames and comparison with MFCC features in a Kaldi environment is planned for the nearest future. 

\subsection{Dataset}

The GoogleSpeechCommands \cite{GoogleSpeechCommands} dataset was used for training and testing. Dataset includes audio fragments of 30 different commands, spoken in noisy conditions. The choice of this dataset was mainly determined by the relative simplicity to test it for both VAE and MFCC features: it is fixed-length audio while containing one of the 30 commands in a single audio. One of the other main reasons was that we have mentioned earlier, that VAE has a great potential for smoothing audio and performing general noise reduction, and it was interesting to test the system on noisy audio. Training was performed on 46619 samples, testing on 11635 samples from the dataset. Audio data had the same format as LibriSpeech: mono-channel wav, sampling rate 16 kHz, signed PCM little endian with bit rate 256 kB/s and bit depth 16.

VAE features were generated in a window 0.3 s with 0.1 overlap. The resulting vectors were then concatenated to serve as a representation of the feature vector of a spoken command.

For the MFCC features an analysis window of length 0.025 s was used with a step of 0.01 s and 26 filterbanks. For one window 13-dimensional cepstrum vector was generated.

\subsection{Architecture}

\begin{figure}[t]
  \centering
  \includegraphics[width=\linewidth]{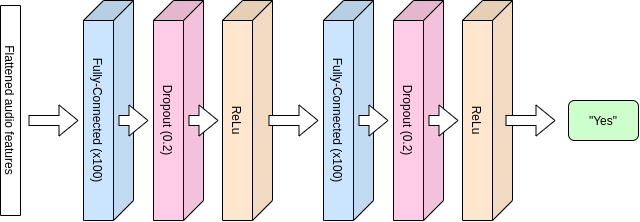}
  \caption{ASR Neural Network structure.}
  \label{fig:asr_nn}
\end{figure}

With this type of dataset it is possible to solve a simple multiclass task, since all the audios are of the same length and one audio only belongs to one class simultaneously. The ASR system architecture was a basic Multi-Layered Perceptron (MLP) of the following structure: 2 fully-connected layers of 100 hidden units with 0.2 dropout rate and ReLU activation followed by a softmax layer (Figure~\ref{fig:asr_nn}).

Although we are solving such a simple ASR task (command recognition), that is quite far away from many practical ASR systems, still solving this task may give us the understanding the potential of the VAE features in comparison to MFCC.

\subsection{Results}

The results of the experiments are shown in the Table 1 below.

\begin{table}[th]
\caption{Speech recognition results}
\label{sample-table}
\vskip 0.15in
\begin{center}
\begin{small}
\begin{sc}
\begin{tabular}{lcc}
\toprule
\textbf{Results} & \textbf{VAE} & \textbf{MFCC} \\
\midrule
Train accuracy & \textbf{0.45} & 0.41 \\ 
Test accuracy & \textbf{0.49} & \textbf{0.49} \\
MLP training time & \textbf{42 s} & 63 s \\
Best epoch & \textbf{12} & 17 \\
Train size & \textbf{204 MB} & 668 MB \\
\bottomrule
\end{tabular}
\end{sc}
\end{small}
\end{center}
\vskip -0.1in
\end{table}

What we can see from the results in the table is that VAE does not concede to the MFCC features, while generally taking up less space. That means that being 3x smaller, VAE features carry approximately the same amount of acoustic information as the MFCC features. This can be very useful in the cases when the audio information has to be stored somewhere (e.g. etalon voice samples, database of speech segments). Furthermore, when the input to the neural network is smaller, as in the case of VAE features in comparison to MFCC, then the amount of parameters in a network also becomes smaller and this leads to faster convergence of the neural network (as it is also highlighted in the results table).

\section{Conclusions}
\label{conclusions}

This research has shown that convolutional VAEs are capable of reconstructing audio data in the form of spectrograms. They succeed at representing fixed-length audio data fragments for the ASR tasks. As a result, convolutional VAEs feature extracting have shown relatively good results in a recognition task on a noisy dataset, in comparison to traditional MFCC features. Finally, the embeddings, generated by VAE occupy 3 times less space in comparison to MFCC, despite being as informative. The resulting algorithms, scripts and models are available on GitHub for any researcher that would be interested.

The further work that has to be done is to test the 13-dimensional latent features of the VAE as an input for a proper ASR system (e.g. Kaldi, DeepSpeech).It is planned to carry out the experiment using Kaldi in the nearest future. This will be done for English and Russian languages, for the Voxforge dataset.

\bibliographystyle{IEEEtran}

\begin{thebibliography}{9}
   \bibitem[1]{Amodei2016}
    Amodei, Dario et al
    \newblock “Deep Speech 2 : End-to-End Speech Recognition in English and Mandarin,” ICML, 2016.

   \bibitem[2]{Dai2017}
    Dai, Bin et al
    \newblock “Hidden Talents of the Variational Autoencoder,” 2017.
    
   \bibitem[3]{Oord2017}
    Oord, Aäron van den et al
    \newblock “Neural Discrete Representation Learning,” NIPS, 2017.
    
   \bibitem[4]{Oord2018}
    Oord, Aäron van den et al
    \newblock “WaveNet: A Generative Model for Raw Audio,” SSW, 2016.

   \bibitem[5]{Zhu2018}
    Zhu, Ziwei et al
    \newblock “Siamese Recurrent Auto-Encoder Representation for Query-by-Example Spoken Term Detection,” Interspeech, 2018.

   \bibitem[6]{Milde2018}
    Milde, Benjamin and Christian Biemann
    \newblock “Unspeech: Unsupervised Speech Context Embeddings,” Interspeech, 2018.

   \bibitem[7]{Chung2018}
    Chung, Yu-An and James R. Glass
    \newblock “Speech2Vec: A Sequence-to-Sequence Framework for Learning Word Embeddings from Speech,” Interspeech, 2018.
    
   \bibitem[8]{LibriSpeech}
    LibriSpeech
    \newblock http://www.openslr.org/12/.

   \bibitem[9]{GoogleSpeechCommands}
    GoogleSpeechCommands
    \newblock ai.googleblog.com/2017/08/launching-speech-commands-dataset.html.

   \bibitem[10]{Audio_vae}
    Audio\_vae
    \newblock {https://github.com/nsu-ai-team/audio\_vae.}

\end{thebibliography}

\end{document}